\documentclass[aps,%
final,%
notitlepage,%
oneside,%
twocolumn,
nobibnotes,%
nofootinbib,%
superscriptaddress,%
showpacs,%
centertags,
floatfix]%
{revtex4}
\usepackage[dvips]{graphicx}
\usepackage{epsfig,graphics,graphicx}

\begin{document}

\title{Towards single-cycle squeezing in chirped quasi-phase-matched optical parametric down-conversion}

\author{D.~B.~Horoshko}
\affiliation{Laboratoire PhLAM, Universit\'e Lille 1, 59655 Villeneuve d'Ascq, France}
\affiliation{B.~I.~Stepanov Institute of Physics, NASB, Nezavisimosti Ave.~68, Minsk 220072 Belarus}%

\author{M.~I.~Kolobov}
\affiliation{Laboratoire PhLAM, Universit\'e Lille 1, 59655 Villeneuve d'Ascq, France}

\date{\today}

\begin{abstract}
We propose a method for generation of a single-cycle squeezed light by parametric down-conversion in a chirped
quasi-phase-matched nonlinear crystal. We find an exact quantum solution for this process valid for an arbitrary parametric gain,
and discover an ultrabroadband squeezing in the down-converted light with flat squeezing spectrum comprising the full optical octave. We describe a scheme for observation of this kind of squeezing using second-harmonic generation as an ultrafast correlator.
\end{abstract}

\maketitle

Squeezed light is one of the central phenomena in the modern quantum optics, being on the one hand a meso- or even macroscopic object with essentially quantum properties, and on the other hand a valuable resource for metrology and for quantum information processing. Both the degree of squeezing and the squeezing bandwidth are important for potential applications. At present, the values of squeezing as high as 11,5 dB  in a band of 100 MHz~\cite{Schnabel10} and 0.3 dB in a band of 2 GHz~\cite{Schnabel12} have been observed in CW regime. Current experiments with pulsed squeezed light reach a bandwidth in the THz~\cite{Leuchs98,Grangier04,Leuchs10,Fabre12} and even tens of THz \cite{Leuchs09} range. In this Letter, we describe a technique for generating a squeezed light with ultimate possible squeezing bandwidth comprising all optical spectrum, 250 THz in a realistic example, the degree of squeezing being almost constant within this huge bandwidth.

The proposed technique is based on the process of parametric down conversion (PDC) of light in a quasi-phase-matched (QPM) periodically poled nonlinear crystal with linear chirp of the poling frequency, known to be able to amplify or generate ultrabroadband optical fields. Linearly chirped QPM crystals were shown to produce single-cycle biphotons in the low-gain regime of PDC~\cite{Harris07,Harris10,Sergienko08} and to provide amplification of classical pulses with high constant gain over a broad bandwidth~\cite{Fejer08a,Fejer08b,Keller10}. In particular, in Ref.~\cite{Keller10} a parametric amplification in a mid-IR of pulses as short as 75 fs with parametric gain over 40 dB has been demonstrated. Here we present a quantum theory of PDC in QPM media in the high-gain regime and demonstrate that this process can be used for generating utrabroadband squeezed light with squeezing spectrum comprising the full optical octave.

We consider the process of collinear non-degenerate type-I PDC, where one photon of the pump wave with frequency $\omega_p$ is annihilated to create one photon of the signal wave with frequency $\omega_0+\Omega$ and one photon of the idler wave with frequency $\omega_0-\Omega$, where $\omega_0=\omega_p/2$. The widths of the signal and idler spectra in PDC are limited by the phase-matching condition. In a QPM crystal with linear chirp this condition can be satisfied for a broad band of frequencies at different spatial positions~\cite{Harris07}.

We describe the down-converted field in the Fourier domain by photon annihilation operators $a(\Omega,z)$, corresponding to annihilation of a photon with frequency $\omega_0+\Omega$ at point $z$. The slowly varying operators $b(\Omega,z)$ are defined by the relations $a(\Omega,z)=b(\Omega,z)\exp[ik(\Omega)z]$, where $k(\Omega)$ is the wave vector at frequency $\omega_0+\Omega$. For a fixed value of $\Omega>0$ we have two independent quantum operators, $b(\Omega,z)$ and $b^\dagger(-\Omega,z)$. In a QPM crystal with the spatial frequency $K=K_0-\zeta z$, where $\zeta$ is the chirp parameter, these two operators are coupled via~\cite{Harris07}
\begin{eqnarray}\label{systemb1}
     \frac{\partial b(\Omega,z)}{\partial z}&=&i\kappa b^\dagger(-\Omega,z) e^{i\zeta z^2/2+i\Delta(\Omega)z},
\end{eqnarray}
where $\kappa$ is the nonlinear coupling coefficient and $\Delta(\Omega)=k_p-[k(\Omega)+k(-\Omega)+K_0]$ is the phase mismatch, $k_p$ being the wave vector of the plane monochromatic pump wave. Note, that the phase mismatch is an even function of the detuning $\Omega$, as usual for a type-I PDC, which implies an important symmetry of Eq.~(\ref{systemb1}) and its solution.

To find the solution of Eq.~(\ref{systemb1}) we rewrite it in a more suitable form, introducing the operators $\tilde b(\Omega,z)$ by
\begin{equation}\label{btilde}
     a(\Omega,z)=\tilde b(\Omega,z)e^{ik(\Omega)z} e^{\frac{i}2(\Delta(\Omega)z+
     \zeta z^2/2+\varphi+\pi/2)},
\end{equation}
where $\varphi={\rm arg}(\kappa)$ is the phase of the pump field.
Next, we introduce a new variable $x=\Delta(\Omega)/\sqrt{\zeta}+z\sqrt{\zeta}$. In the variables $(\Omega,x)$ Eq.~(\ref{systemb1}) becomes
\begin{eqnarray}\label{systembt1}
     \frac{\partial\tilde b(\Omega,x)}{\partial x}+\frac{i}2x\tilde b(\Omega,x)=\sigma\tilde b^\dagger(-\Omega,x),
\end{eqnarray}
where $\sigma=|\kappa|/\sqrt{\zeta}$ is the new nonlinear coupling parameter. The system of two linear first-order differential equations, Eq.~(\ref{systembt1}) and its Hermitian conjugate with sign inversion for $\Omega$, is equivalent to one second-order equation
\begin{equation}\label{second}
     \frac{\partial^2\tilde b(\Omega,x)}{\partial x^2}+\left(\frac14 x^2+\frac{i}2-\sigma^2\right)\tilde b(\Omega,x)=0,
\end{equation}
having solutions in the class of parabolic cylinder functions~\cite{Abramowitz}. Let us denote two linearly independent solutions of Eq.~(\ref{second}) as $\phi_1(x)$ and $\phi_2(x)$ with constant Wronskian $W$. These two functions can be chosen among various pairs of special functions of the mentioned class. We introduce the ``reciprocal'' functions  $\tilde\phi_i(x)$, $i=1,2$, by the relation
\begin{equation}\label{reciprocal}
     \frac1{\sigma}\left(\frac{\partial}{\partial x}+\frac{i}2x\right) \phi_i(x)=\tilde\phi_i(x).
\end{equation}
By construction, the pairs $(\phi_i(x),\tilde\phi_i(x))$ are solutions of the system created by Eq.~(\ref{systembt1}) and its Hermitian conjugate with sign-inverted $\Omega$. The general solution of this system with boundary conditions at $x=x_0$ is written as
\begin{eqnarray}\nonumber
     \tilde b(\Omega,x)&=&\frac{\sigma}W\left\{\left| \begin{array}{cc}\phi_1(x)&\phi_2(x)\\\tilde\phi_1(x_0)&\tilde\phi_2(x_0)\end{array} \right|\tilde b(\Omega,x_0)\right.\\\label{solution1}
     &&\left.-\left| \begin{array}{cc}\phi_1(x)&\phi_2(x)\\\phi_1(x_0)&\phi_2(x_0)\end{array} \right|\tilde b^\dagger(-\Omega,x_0)\right\},\\\nonumber
     \tilde b^\dagger(-\Omega,x)&=&\frac{\sigma}W\left\{-\left| \begin{array}{cc}\tilde\phi_1(x)&\tilde\phi_2(x)\\\phi_1(x_0)&\phi_2(x_0)\end{array} \right|\tilde b^\dagger(-\Omega,x_0)\right.\\\nonumber
     &&\left.+\left| \begin{array}{cc}\tilde\phi_1(x)&\tilde\phi_2(x)\\\tilde\phi_1(x_0)&\tilde\phi_2(x_0)\end{array} \right|\tilde b(\Omega,x_0)\right\}.
\end{eqnarray}
Equations (\ref{solution1}) represent the Bogoliubov transformation of the field operators, which is known to produce multimode squeezed states of light~\cite{Kolobov1999}.

For practical calculations we choose the functions $\phi_1(x)$ and $\phi_2(x)$ from the family of Whittaker parabolic cylinder functions~\cite{Abramowitz}:
\begin{eqnarray}\label{parabolic1}
     \phi_1(x)&=&D_{i\nu}(xe^{i\pi/4}),\\\nonumber
     \phi_2(x)&=&D_{-1-i\nu}(-xe^{-i\pi/4}),
\end{eqnarray}
with the corresponding reciprocal functions
\begin{eqnarray}\label{parabolic3}
\tilde\phi_1(x)&=&\nu^{1/2}e^{i3\pi/4}D_{i\nu-1}(xe^{i\pi/4}),\\\nonumber
\tilde\phi_2(x)&=&\nu^{-1/2}e^{-i\pi/4}D_{-i\nu}(-xe^{-i\pi/4}),
\end{eqnarray}
where $\nu=\sigma^2$. The Wronskian of $\phi_1(x)$ and $\phi_2(x)$ is $W=e^{-i\pi/4+\pi\nu/2}$.

Using these functions as the basis, we find from Eqs.~(\ref{solution1}) the transformation of the field operators $\tilde b(\Omega,z)$ from the crystal input at $z=0$ to its output at $z=L$:
\begin{eqnarray}\label{transf1}
     \tilde b(\Omega,L)&=&A(\Omega)\tilde b(\Omega,0) + B(\Omega)\tilde b^\dagger(-\Omega,0),\\\nonumber
     \tilde b^\dagger(-\Omega,L)&=&\tilde A(\Omega)\tilde b^\dagger(-\Omega,0) +\tilde B(\Omega)\tilde b(\Omega,0),
\end{eqnarray}
where
\begin{eqnarray}\label{U}
     A(\Omega)&=&\left[ D_{i\nu}(x_Le^{i\pi/4})D_{-i\nu}(-x_0e^{-i\pi/4})\right.\\\nonumber &&\left.+\nu D_{-1-i\nu}(-x_Le^{-i\pi/4})D_{i\nu-1}(x_0e^{i\pi/4})\right]e^{-\pi\nu/2},\\\nonumber
     B(\Omega)&=&\sigma e^{i\pi/4}\left[ D_{-1-i\nu}(-x_Le^{-i\pi/4})D_{i\nu}(x_0e^{i\pi/4})\right.\\\nonumber &&\left.- D_{i\nu}(x_Le^{i\pi/4})D_{-1-i\nu}(-x_0e^{-i\pi/4})\right]e^{-\pi\nu/2},
\end{eqnarray}
while $\tilde A(\Omega)$ and $\tilde B(\Omega)$ are obtained from $A(\Omega)$ and $B(\Omega)$ by mutual exchange of $\phi_1(x)$ and $\phi_2(x)$ with their reciprocal functions.

Frequency enters into Eqs.~(\ref{U}) only via the dependence of $x_L=\Delta(\Omega)/\sqrt{\zeta}+L\sqrt{\zeta}$ and $x_0=\Delta(\Omega)/\sqrt{\zeta}$ of the phase mismatch, meaning that $A(\Omega)$, $B(\Omega)$, $\tilde A(\Omega)$ and $\tilde B(\Omega)$ are even functions of the detuning $\Omega$. It will be shown elsewhere~\cite{FullPaper} that $\tilde A^{}(\Omega)=A^*(\Omega)$, $\tilde B^{}(\Omega)=B^*(\Omega)$, though in general $\tilde\phi_n(x)\ne\phi^*_n(x)$.

In Eqs.~(\ref{U}) we recognize the Green's functions for the signal and the idler waves in a chirped QPM crystal, obtained by a purely classical treatment of the problem~\cite{Fejer08b}. In our approach Eqs.~(\ref{transf1}) describe a unitary transformation of the field operators, realized in the nonlinear medium. The unitarity of the transformation requires that $|A(\Omega)|^2-|B(\Omega)|^2=1$, and $A(\Omega)/B(\Omega)=A(-\Omega)/B(-\Omega)$. The first relation can be proven~\cite{FullPaper} with the help of the recurrence relations for the parabolic cylinder functions~\cite{Abramowitz}, while the second relation follows from the evenness of the functions $A(\Omega)$ and $B(\Omega)$.

Using Eqs.~(\ref{btilde}) and (\ref{transf1}) we obtain the transformation of the field operator $a(\Omega,z)$ in the form of one equation:
\begin{eqnarray}\label{transf}
     a(\Omega,L)&=&U(\Omega)a(\Omega,0) +V(\Omega)a^\dagger(-\Omega,0),
\end{eqnarray}
where
\begin{eqnarray}\label{Ufull}
     U(\Omega)&=&A(\Omega)e^{ik(\Omega)L} e^{\frac{i}2(\Delta(\Omega)L+\zeta L^2/2)},\\\nonumber
     V(\Omega)&=&i B(\Omega)e^{ik(\Omega)L} e^{\frac{i}2(\Delta(\Omega)L+\zeta L^2/2)+i\varphi}.
\end{eqnarray}
Note that these two functions of $\Omega$ are not even anymore, because, due to dispersion, the signal and the idler waves acquire different values of phase $k(\Omega)L$ when passing through the crystal. The properties of the functions $U(\Omega)$ and $V(\Omega)$, required by the unitarity of Eq.~(\ref{transf}), follow directly from that of functions $A(\Omega)$ and $B(\Omega)$.

With the help of asymptotic properties of the parabolic cylinder functions, various limiting forms of Eqs.~(\ref{U}) can be explored. In the case of low gain, $\nu\ll 1$, we can take the limit $\nu\rightarrow 0$, and using the relations~\cite{Abramowitz} $D_{0}(x)=\exp[-x^2/4]$, $D_{-1}(x)=\sqrt{\pi/2}\exp[x^2/4]\left(1-i{\rm erfi}[-ix/\sqrt{2}]\right)$, we obtain
\begin{eqnarray}\label{Ulow}
    A(\Omega)&\rightarrow&e^{-i(x_L^2-x_0^2)/4},\\\label{Vlow}
    B(\Omega)&\rightarrow&ie^{i\pi/4}\sigma \sqrt{\frac{\pi}2}e^{-i(x_L^2+x_0^2)/4}\\\nonumber
&&\times\left\{{\rm erfi}\left[\frac{(1+i)x_0}2\right]-{\rm erfi}\left[\frac{(1+i)x_L}2\right]\right\},
\end{eqnarray}
which, together with the phase factor from Eqs.~(\ref{Ufull}) give the transformation, obtained in the low-gain regime as a perturbative solution of the initial Eq.~(\ref{systemb1})~\cite{Harris07}.

\begin{figure}[h]
\center{\includegraphics[width=\linewidth]{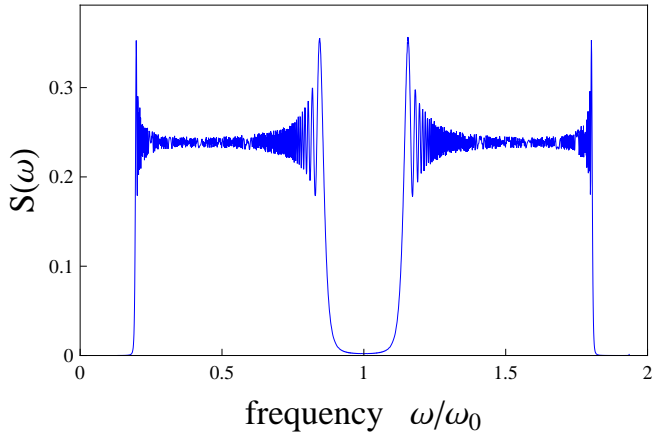}}
\caption{Optical spectrum of the ultrabroadband squeezed light for the 2 cm crystal of LiNbO$_3$, pumped at 0.42 $\mu$m. The spectrum is symmetric with respect to the frequency $\omega_0$, as implied by the evenness of $|V(\Omega)|$.}
\label{SpectrumHighGain}
\end{figure}

Using the solution given by Eq.(\ref{transf}), it is straightforward to find the optical spectrum $S(\omega)$ of the PDC light at the
output of the crystal, defined by the relation $\langle a^{\dagger}(\Omega,L)a(\Omega',L)\rangle=S(\omega_0+\Omega)\delta(\Omega-\Omega')$. Employing the canonical commutation relations,
$[\tilde b(\Omega,0),\tilde b^{\dagger}(\Omega',0)]=\frac1{2\pi}\delta(\Omega-\Omega')$, we obtain
\begin{equation}\label{S}
     S(\omega_0+\Omega)=\frac1{2\pi}\left|V(\Omega)\right|^2.
\end{equation}
This spectrum is shown in Fig.~\ref{SpectrumHighGain} for the 2 cm crystal of LiNbO$_3$, pumped at 0.42 $\mu$m and quasi-phase-matched to produce down converted light from 0.46 to 0.75 $\mu$m, as in Ref.~\cite{Harris07}. The refractive index is obtained from the Sellmeier equation for extraordinary wave in LiNbO$_3$. The difference from Ref.~\cite{Harris07} is in the pump intensity which is 4 orders of magnitude higher, corresponding to $\nu=0.146$.

As well known~\cite{Kolobov1999}, the transformation given by Eq.~(\ref{transf}) generates broadband quadrature squeezing in the PDC light. For each pair of modes with opposite detunings we construct two quadrature operators as~\cite{Kolobov1999}
\begin{eqnarray}\label{quad1}
     X_1(\Omega,L)&=&b(\Omega,L)e^{i\psi(\Omega,L)} +b^{\dagger}(-\Omega,L)e^{-i\psi(\Omega,L)},\\
     \nonumber
     X_2(\Omega,L)&=&-i\left[b(\Omega,L)e^{i\psi(\Omega,L)} -b^{\dagger}(-\Omega,L)e^{-i\psi(\Omega,L)}\right],
\end{eqnarray}
where the angle of squeezing $\psi(\Omega,L)=\frac12{\rm \arg}[U(\Omega)V(-\Omega)]$ determines the orientation of the squeezing ellipse. In terms of these quadratures the transformation Eq.~(\ref{transf}) can be rewritten as
\begin{equation}\label{X}
     X_{\mu}(\Omega,L)=e^{ik_-(\Omega)L}\exp[\pm r(\Omega)]X_{\mu}(\Omega,0),
\end{equation}
where the upper (lower) sign corresponds to $\mu=1$ ($\mu=2$), $\exp[\pm r(\Omega)]=|U(\Omega)|\pm |V(\Omega)|$, and $k_-(\Omega)=[k(\Omega)-k(-\Omega)]/2$. It follows from Eq.~(\ref{X}) that the quadrature $X_2(\Omega,L)$ is squeezed below the standard quantum limit, while the conjugate quadrature $X_1(\Omega,L)$ is stretched above that limit.

\begin{figure}[h]
\center{\includegraphics[width=\linewidth]{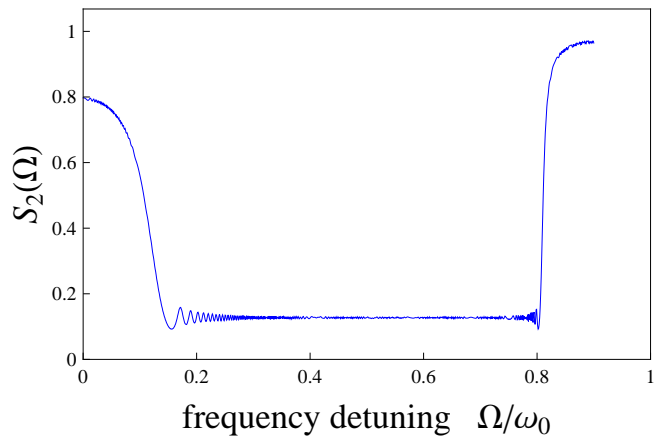}}
\caption{Spectrum of squeezing in the high-gain regime for the same settings as in Fig.~\ref{SpectrumHighGain}, $\nu=0.146$.  }
\label{SqueezingSpectrum}
\end{figure}

The spectra of the quadratures components are defined as follows: $\langle X_{\mu}(\Omega,L)X_{\mu}(\Omega',L)\rangle=S_{\mu}(\Omega)\delta(\Omega+\Omega')$, the spectrum of the squeezed quadrature $X_2$ being known as the spectrum of squeezing $S_{2}(\Omega)=\exp[-2r(\Omega)]$. This spectrum is plotted in Fig.~\ref{SqueezingSpectrum} for the same crystal and experimental settings as in Fig.~\ref{SpectrumHighGain}.

As follows from Fig.~\ref{SqueezingSpectrum}, an octave-broad squeezing can be generated in realistic QPM crystals, widely available today. The degree of squeezing can be estimated by considering the limit of $x\gg|\nu|$ for the parabolic cylinder functions, which results in approximation of the modulus of $U(\Omega)$ within the squeezing band by the Rosenbluth gain factor \cite{Fejer08b,Rosenbluth1973}: $|U(\Omega)|\approx e^{\pi\nu}$, and consequently, $S_2(\Omega)\approx (e^{\pi\nu}-\sqrt{e^{2\pi\nu}-1})^2$. In the low-gain regime $S_2(\Omega)\approx 1$ at all frequencies.

\begin{figure}[h]
\center{\includegraphics[width=\linewidth]{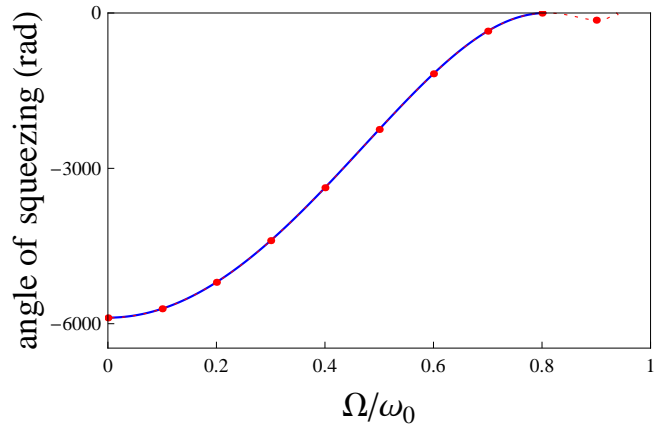}}
\caption{Angle of squeezing $\psi(\Omega,L)$ in the high-gain regime (blue solid line) and the low-gain compensation angle $\theta_0(\Omega)$ (red circles and dotted line). The crystal is the same as in Figs.~\ref{SpectrumHighGain} and \ref{SqueezingSpectrum}. The phase of the pump field $\varphi$ is taken as zero. }
\label{SqueezingAngle}
\end{figure}

The angle of squeezing $\psi(\Omega,L)$ is calculated numerically and shown in Fig.~\ref{SqueezingAngle}. It spans 6000 rad through a bandwidth of $1.5\times10^{15}$Hz, or on average 0.004 rad/GHz. This angle is compared to the compensation angle $\theta_0(\Omega)=\Delta(\Omega)^2/(4\zeta)-[k(\Omega)+k(-\Omega)]L/2$ found for the low-gain regime~\cite{Harris07}. We see these two angles have similar behavior. Therefore, the angle of squeezing can be compensated by the same technique as in Refs.~\cite{Harris07,Harris10}.

In the time domain for coherent field the relation $\langle\delta X_{\mu}(t)\delta X_{\mu}(t')\rangle=\delta(t-t')$, where $\delta X=X-\langle X\rangle$, imposes the standard quantum limit of the quadrature measurements. In the limiting case of a flat squeezing spectrum equal to $e^{-2r}$ from 0 to $\omega_0$ and a constant $\psi(\Omega,L)$, taking a Fourier transform of Eq.(\ref{quad1}) we obtain
\begin{eqnarray}\label{time}
     \langle\delta X_{1}(t)\delta X_{1}(t')\rangle&=&e^{2r}\tilde\delta(t-t'),\\
     \nonumber
     \langle\delta X_{2}(t)\delta X_{2}(t')\rangle&=&e^{-2r}\tilde\delta(t-t'),
\end{eqnarray}
where $\tilde\delta(t)=\frac{\omega_0}{\pi}{\rm sinc}(\omega_0t)$ is a delta-like function with the width of the order of an optical period at the carrier frequency $\omega_0$. Equations (\ref{time}) describe an ultimate limit of squeezing in temporal domain with the characteristic time as short as one optical period, the phenomenon which we call ``single-cycle squeezing''. This kind of squeezing would have numerous potential applications, for example, in the quantum metrology of ultrashort optical processes. The squeezing parameter $r$ will be determined by the available parametric gain, which can be very high for nanosecond~\cite{Fejer08a} or femtosecond~\cite{Keller10} optical parametric amplification.

The ultrabroadband nature of the generated squeezed light can be observed using second-harmonic generation as an ultrafast correlator, similar to the experiments in the low-gain regime~\cite{Silberberg05a,Silberberg05b,URen09}. When the PDC light is directed to a thin crystal allowing the second-harmonic generation, the field of second harmonic can be written as
$
    a_2(t)=a_{20}(t)+\epsilon a^2(t),
$
where $a_{20}(t)$ is the vacuum field of the second harmonic at the input,  $a(t)$ is the field of the PDC light, and $\epsilon$ is a small parameter, determining the efficiency of the process. The spectrum of the second harmonic $S_2(\omega)$, determined by the relation $\langle a_2^{\dagger}(\omega)a_2(\omega')\rangle= S_{SH}(\omega)\delta(\omega-\omega')$, reads as:
\begin{eqnarray}\label{a2s}
     S_{SH}(\omega)&=&\delta(\omega-\omega_p) \left|\frac{\epsilon}{2\pi} \int_{-\omega_0}^{\omega_0}U_c(\Omega)V_c(-\Omega)d\Omega\right|^2\\\nonumber
     &&+\left(\frac{\epsilon}{2\pi}\right)^2 2\int_{-\omega_0}^{\omega_0} \left|V(\Omega)V(\omega-\omega_p-\Omega)\right|^2d\Omega.
\end{eqnarray}
where $U_c(\Omega)=U(\Omega)e^{-i\theta(\Omega)}$, $V_c(\Omega)=V(\Omega)e^{-i\theta(\Omega)}$ are the transformation coefficients including the effect of compensation by a phase shift $\theta(\Omega)$.
The spectrum Eq.(\ref{a2s}) contains two components: the coherent one, sharply peaked at the pump frequency, and the incoherent one, spectrally very broad. The field of the coherent component in the photon-flux units reads as
\begin{equation}\label{Ecoh}
    E_{coh}=\frac{\epsilon}{\pi} \int_0^{\omega_0} |U(\Omega)V(\Omega)| e^{i[2\psi(\Omega,L)-\theta(\Omega)-\theta(-\Omega)]}d\Omega.
\end{equation}

\begin{figure}[!h]
\begin{tabular}{cc}
   \includegraphics[width=0.45\linewidth]{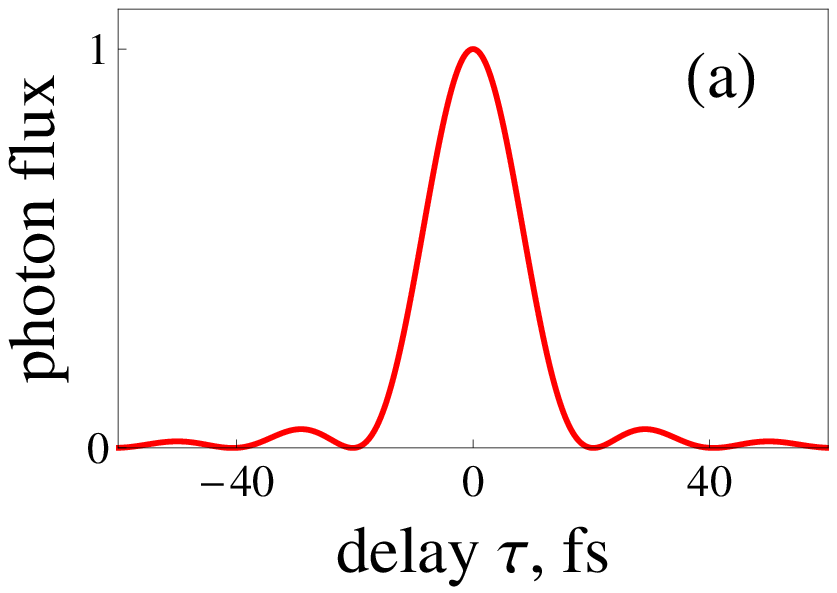} &
   \includegraphics[width=0.47\linewidth]{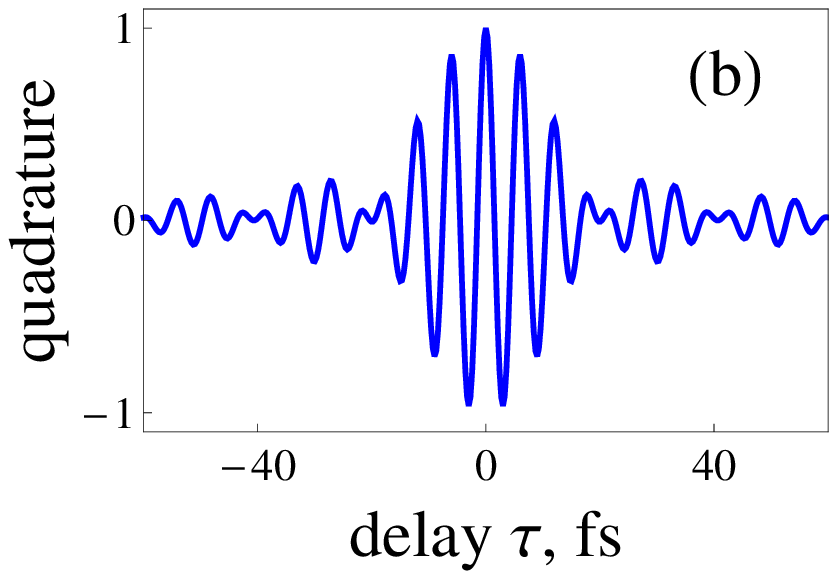} \\
   \includegraphics[width=0.45\linewidth]{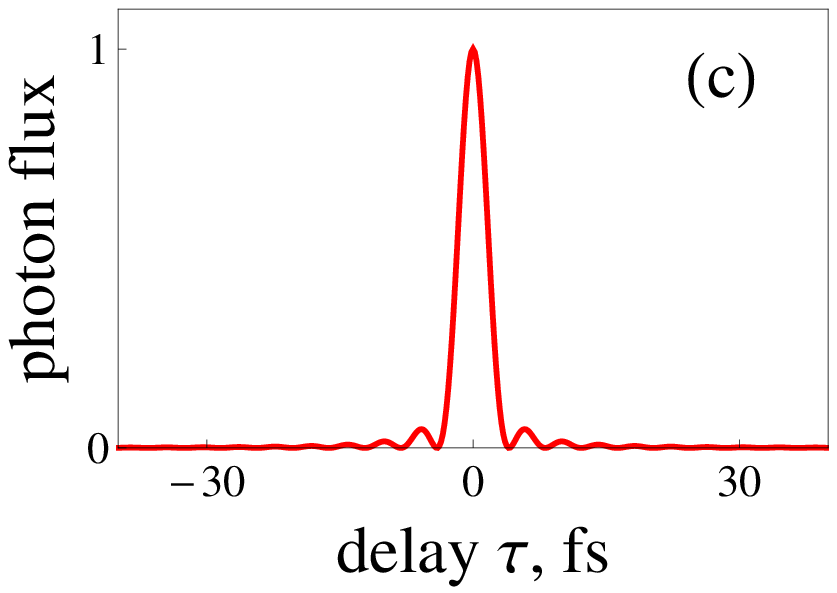} &
   \includegraphics[width=0.47\linewidth]{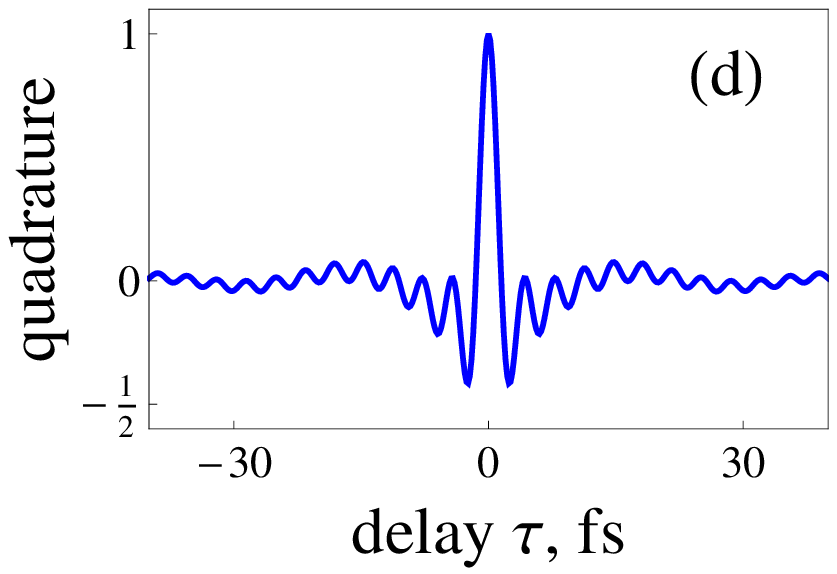}
\end{tabular}
\caption{Normalized photon flux $\Phi(\tau)$ (a,c) and normalized amplitude quadrature $X(\tau)$ (b,d) for the coherent component of the second harmonic as functions of the delay time for the cases of low chirp (a,b), $\zeta=1.14\times10^{7}m^{-2}$, and high chirp (c,d), $\zeta=5.64\times10^{7}m^{-2}$.}
\label{HarmonicIntensity}
\end{figure}

If the compensation angle matches exactly the squeezing angle $\theta(\Omega)=\psi(\Omega,L)$, the total photon flux of the coherent component of the second harmonic reaches its maximum value. If, in addition to compensation, a small delay $\tau$ is introduced into the signal wave, $\theta(\Omega)=\psi(\Omega,L)+\tau\Omega$, $\Omega>0$, then the coherent component disappears as soon as $\tau$ exceeds the inverse of the squeezing spectrum width $2\pi/\Delta\Omega$, which for the crystal discussed above is 4.1 fs. Approximating the squeezing spectrum by a rectangle within the squeezing band, we obtain:
\begin{eqnarray}\label{Ecoh2}
     E_{coh}(\tau)&\approx&\frac{\epsilon}{\pi}U_0V_0\Delta\Omega e^{-i\Omega_s\tau}{\rm sinc}\left[\Delta\Omega\tau/2\right],
\end{eqnarray}
where $\Omega_s$ is the central frequency of the signal, while $U_0$ and $V_0$ are the values of $\left|U(\Omega)\right|$ and $\left|V(\Omega)\right|$ within the squeezing band.

The measurement of the dependence of the coherent component photon flux $\Phi(\tau)=|E_{coh}(\tau)|^2$ on the small delay $\tau$ (Fig.~\ref{HarmonicIntensity}) would be, thus, a detection of the ultrabroad bandwidth of the squeezed light. Homodyne detection of the second harmonic field with the PDC pump as a local oscillator \cite{Harris07} will allow one to observe the amplitude quadrature $X(\tau)=2{\rm Re}\{e^{-i\varphi}E_{coh}(\tau)\}$, oscillating with delay at the optical frequency with an envelope determined by the correlation time, and thus providing a direct comparison of the correlation time to the optical cycle. In Fig.~\ref{HarmonicIntensity} we show the photon flux and the quadrature for a low-chirp crystal with a spectrum 5 times narrower than that considered above, where the correlation time exceeds significantly the optical period, and the same quantities for the high-chirp crystal, where these times are of the same order.

Ultrabroaband multimode structure of single-cycle squeezed light can be also investigated using the methods developed for studying multimode quantum frequency combs~\cite{Fabre12}, squeezed quantum pulses~\cite{Opatrny2002}, and quantum fluctuations in optical solitons~\cite{Spalter1998}. We shall address this topic elsewhere~\cite{FullPaper}.

In conclusion, we have presented a method for generation of a single-cycle squeezed light as an ultimate limit of ultrabroadband squeezed light with squeezing comprising the full optical spectrum. Our scheme is based on parametric down-conversion in a linearly chirped quasi-phase-matched nonlinear crystal. Our proposal is within the reach of current technologies related to femtosecond optical parametric amplification in such media.

We are grateful to Steve Harris for valuable discussions. This work was supported by region Nord-Pas-de-Calais (France) via project ``Campus Intelligence Ambiante''.


\begin{thebibliography}{00}
\bibitem{Schnabel10} M.~Mehmet \textit{et al.}, Phys.~Rev.~A {\bf 81}, 013814 (2010).
\bibitem{Schnabel12} S.~Ast \textit{et al.}, Opt.~Lett.~\textbf{37}, 2367 (2012).
\bibitem{Leuchs98} S.~Spaelter, N.~Korolkova, F.~Koenig, A.~Sizmann, and G.~Leuchs, Phys.~Rev.~Lett.~\textbf{81}, 786 (1998).
\bibitem{Grangier04} J.~Wenger, R.~Tualle-Brouri, and P.~Grangier, Opt.~Lett.~\textbf{29}, 1267 (2004).
\bibitem{Leuchs10} I.~N.~Agafonov, M.~V.~Chekhova, and G.~Leuchs, Phys.~Rev.~A \textbf{82}, 011801 (2010).
\bibitem{Fabre12} O.~Pinel, P.~Jian, R.~Medeiros de Araujo, J.~Feng, B.~Chalopin,
C.~Fabre, and N.~Treps, Phys.~Rev.~Lett.~\textbf{108}, 083601 (2012).
\bibitem{Leuchs09} T.~Iskhakov, M.~V.~Chekhova, and G.~Leuchs, Phys.~Rev.~Lett.~\textbf{102}, 183602 (2009).
\bibitem{Harris07} S.~E.~Harris, Phys.~Rev.~Lett.~\textbf{98}, 063602 (2007).
\bibitem{Harris10} S.~Sensarn, G.~Y.~Yin, and S.~E.~Harris, Phys.~Rev.~Lett.~\textbf{104}, 253602 (2010).
\bibitem{Sergienko08} M.~B.~Nasr~\textit{et al.}, Phys.~Rev.~Lett.~\textbf{100}, 183601 (2008).
\bibitem{Fejer08a} M.~Charbonneau-Lefort, B.~Afeyan, and M.~M.~Fejer, J.~Opt.~Soc.~Am.~B \textbf{25}, 463 (2008).
\bibitem{Fejer08b} M.~Charbonneau-Lefort, B.~Afeyan, and M.~M.~Fejer, J.~Opt.~Soc.~Am.~B \textbf{25}, 680 (2008).
\bibitem{Keller10} C.~Heese, C.~R.~Phillips, L.~Gallmann, M.~M.~Fejer, and U.~Keller, Opt.~Lett.~{\bf 35}, 2340 (2010).
\bibitem{Abramowitz} M.~Abramowitz and I.~A.~Stegun, Eds.~Handbook of Mathematical Functions (National Bureau of Standards, 1972).
\bibitem{Kolobov1999} M.~I.~Kolobov, Rev.~Mod.~Phys.~\textbf{71}, 1539 (1999).
\bibitem{FullPaper} D.~B.~Horoshko and M.I. Kolobov, to be published.
\bibitem{Rosenbluth1973} M.~N.~Rosenbluth, R.~B.~White, and C.~S.~Liu, Phys.~Rev.~Lett.~\textbf{31}, 1190 (1973).
\bibitem{Silberberg05a} B.~Dayan, A.~Pe'er, A.~A.~Friesem, and Y.~Silberberg, Phys.~Rev.~Lett.~\textbf{94}, 043602 (2005).
\bibitem{Silberberg05b} A.~Pe'er, B.~Dayan, A.~A.~Friesem, and Y.~Silberberg, Phys.~Rev.~Lett.~\textbf{94}, 073601 (2005).
\bibitem{URen09} K.~A.~O'Donnell and A.~B.~U'Ren, Phys.~Rev.~Lett.~\textbf{103}, 123602 (2009).
\bibitem{Opatrny2002} T.~Opatrn\'{y}, N.~Korolkova, and G.~Leuchs, Phys.~Rev.~A~\textbf{66}, 053813 (2002).
\bibitem{Spalter1998} S.~Sp\"{a}lter \textit{et al}, Phys.~Rev.~Lett.~\textbf{81}, 786 (1998).

\end{thebibliography}
\end{document}